# Tuning Spontaneous Emission versus Förster Energy Transfer in Biological Systems by Manipulating the Density of Photonic States

Christian Blum[1], Willem L. Vos[2,3], Vinod Subramaniam[1]

[1] Biophysical Engineering Group, Faculty of Science and Technology
and MESA+ Institute for Nanotechnology, University of Twente, P.O. Box 217, 7500 AE Enschede, The Netherlands.

[2] Complex Photonic Systems (COPS), Faculty of Science and Technology
and MESA+ Institute for Nanotechnology, University of Twente, P.O. Box 217, 7500 AE Enschede, The Netherlands.

[3] Center for Nanophotonics, FOM Institute AMOLF, Kruislaan 407, 1098 SJ Amsterdam, The Netherlands.

ABSTRACT:

We theoretically discuss how to tune the competition between Förster transfer and spontaneous emission in a continuous and nondestructive fashion. The proposed approach is especially suitable for delicate biological systems like light harvesting complexes and fluorescent protein oligomers. We demonstrate that the manipulation of the density of photonic states at the emission frequency of the energy donor results in a change of the quantum efficiencies of the competing energy transfer and spontaneous emission processes. This change will be manifested in a modification of the donor and acceptor emission intensities. Thus, by controlling the local density of photonic states Förster coupled systems can be manipulated and analyzed without the need to physically separate donor and acceptor chromophores for individual analysis, which is of interest, for example, for oligomeric reef coral fluorescent proteins.

## I. INTRODUCTION

It is well known that an excited molecule can either spontaneously emit a photon, transfer energy by resonant energy transfer to another molecule, or deactivate by radiationless processes [1]. In biology and life sciences the competition between spontaneous emission and energy transfer is of particular interest. Fluorescence resonance energy transfer plays a key role in biological processes like photosynthesis [2-5]. Moreover the emission and interaction of fluorescent probes like fluorescent proteins are an indispensable tool to reveal protein interactions [6-8], conformational dynamics [9,10] or to measure distances at the nanometer scale [11,12].

The usual approach to study or modify the competition between resonant energy transfer and spontaneous emission is to chemically modify the involved molecules or to substitute them with others. This chemical tuning is a drastic interference with the system: it takes place in discrete steps and is irreversible. Exciting opportunities arise if the tuning could be achieved without discrete chemical alterations in a continuous and nondestructive fashion. Such tuning could find a number of interesting applications. New ways to analyze sensitive complex Förster-coupled biological systems like light harvesting complexes or fluorescent protein oligomers would arise. Such a method also has the potential to make the analysis of biological processes illuminated by the interaction of coupled chromophores faster and more efficient.

In this paper, we propose a new approach to tune the competition between resonant energy transfer and spontaneous emission by manipulating the local photonic environment, described by the so-called local density of states, of the target system. The desired tuning of the local density of states can be achieved by placing the chromophores near a mirror [13], near a dielectric interface [14], or even inside a photonic crystal [15,16]. In this way, we modify the spontaneous emission of the energy donor molecule but not the energy transfer to the acceptor. We predict that the resulting changes in the quantum efficiencies of the two competing processes can be observed as changes in the intensity ratios of donor and acceptor emissions in a spectrally-resolved intensity measurement.

## II. SPONTANEOUS EMISSION OF LIGHT

It is well-known that spontaneous emission of light is not an immutable property of an excited (bio-) molecule [14,15,17-21]. The characteristics of spontaneously emitted light depend both on the emitter, and on the photonic environment of the emitter. In the weak-coupling limit the radiative decay rate $\gamma_{Rad}$ is described by Fermi's "golden rule" [19]:

$$\gamma_{Rad}(\omega,\vec{r}) = \frac{\pi\omega}{3\hbar\varepsilon_0} D^2 \rho(\omega,\vec{r}) \qquad (1),$$

with $\hbar$ the reduced Planck constant and $\varepsilon_0$ the dielectric permittivity. The emission rate is a function of the optical frequency $\omega$ ($\omega=2\pi c/\lambda$ with the wavelength $\lambda$), the position of the emitter $\vec{r}$, and of the orientation of the dipole moment with respect to the field's polarization. The rate is thus determined by the emitter via its transition dipole moment $D$, and by the environment via the local density of states $\rho(\omega,\vec{r})$. The latter counts the available number of electromagnetic modes in which photons can be emitted at the emission frequency $\omega$ and position $\vec{r}$ of the emitter [20,21].

Clearly, control of the local density of states is a fundamental way to manipulate the radiative decay rate $\gamma_{Rad}$ of an emitter. One method is to place the emitter at a specific distance $d$ from a mirror, as first demonstrated by Drexhage [13]. In this case, the local density of states depends on the frequency, which is contained in a dimensionless parameter ($d/\lambda$) and on the orientation of the transition dipole moment with respect to the mirror. It is also possible to place the emitter at a specific distance $d$ from a dielectric interface, as has been precisely studied in ref. [14]. Both photonic geometries have the advantage that the density of states are known and can be easily calculated, but the variations of the density of states are limited to about a factor of 2. A larger variation of the density of states is achieved in so-called photonic crystals [20,21]. Photonic crystals are intricate periodic dielectric nanostructures, in which the refractive index varies spatially on length scales of the lattice parameter $a$ of the order of the wavelength of light [15,16]. An example is shown in the scanning electron micrograph in Figure 1A. Since the structure consists mostly of air and since the air spheres are topologically connected, there is sufficient space to infiltrate macromolecular emitters such as proteins or protein complexes. Figure 1B shows the density of states calculated for such a crystal that has a backbone with a refractive index of 3.36 (see [21,22]). The graph depicts the density of states normalized to the density of states in a homogeneous medium with the same average refractive index as a function of reduced frequency $a/\lambda$. Clearly the density of states can both be strongly increased as well as decreased compared to a homogeneous medium depending on the lattice parameter $a$ of the photonic crystal, resulting in enhanced or inhibited radiative decay rates of emitters localized inside the photonic crystal. Photonic crystals are hotly pursued because of the prediction of a photonic band gap: a range of wavelengths where no modes can propagate at all, leading to a vanishing of the density of states and to a concomitant infinite lifetime for an embedded emitter [15]. Figure 1B shows a band gap near $a/\lambda=0.8$ but it should be noted that this is not essential for the present study. Recently, the first control of the emission rate of light emitters in inverse opal photonic crystals has been demonstrated, where both longer and shorter lifetimes were observed [23]. At least a factor of 8 variation in emission rate has been observed in photonic crystals similar to the one shown in Figure 1A [24].

(A)    (B)

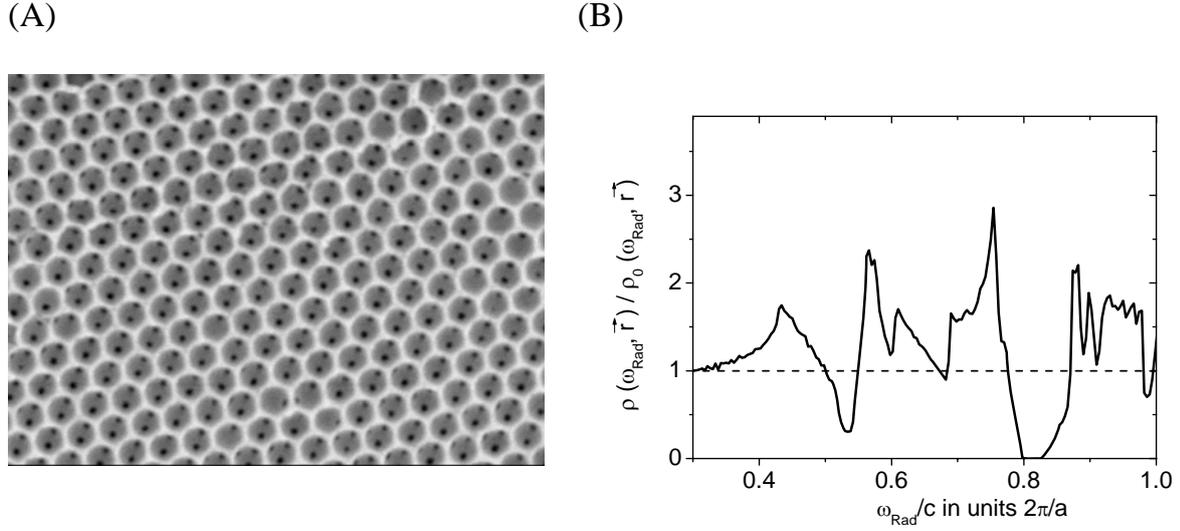

Figure 1 (A). Scanning electron microscope image of the surface of an inverse opal photonic crystal [25,26]. A highly ordered face centered cubic arrangement of air spheres is seen in a highly refractive TiO$_2$ backbone. Here, the sphere radius is 330 nm, and the lattice parameter equals $a$ = 933 nm. The small black holes at the bottom of each air sphere are windows connecting to the next layer of air spheres below. This ensures a three-dimensional connectivity, which is favorable both for band gap formation, and to infiltrate solutions of Förster coupled systems. The backbone material TiO$_2$ is well suited for infiltration of biological systems, since it is inert and non poisonous. (B). Calculated density of states $\rho(\omega_{Rad}, \vec{r})$ normalized to the density of states in a homogeneous medium with the same average refractive index $\rho_0(\omega_{Rad}, \vec{r})$ of an inverse opal photonic crystal with a high refractive index backbone material of 3.36. By changing lattice parameter $a$ of the crystal, the local density of states can be controlled and thus increased or decreased radiative decay rates of emitters inside the photonic crystal can be realized [24].

## III. FÖRSTER RESONANCE ENERGY TRANSFER

Förster transfer or fluorescence resonance energy transfer is the resonant transfer of an energy quantum from an initially excited donor molecule to an acceptor molecule. Fluorescence resonance energy transfer is a non-radiative process, that is, there is no emission and reabsorption of a real photon involved, since the transfer proceeds at distances $d$ much shorter than a wavelength: $d \ll \lambda$. The process is the result of long range dipole-dipole interactions [1,27]. The exquisitely sensitive distance dependence of the transfer rate has led to the widespread use of fluorescence resonance energy transfer as a "molecular ruler" to measure distances in the range of a few nanometers in macromolecules. The expression for the energy transfer rate $\gamma_T$ is:

$$\gamma_T = \frac{1}{\tau_D} \cdot \left(\frac{R_0}{d}\right)^6 \tag{2},$$

where $\tau_D$ is the lifetime of the excited state of the donor molecule in the absence of the acceptor, $d$ is the distance between the acceptor and donor molecules, and $R_0$ is the Förster distance for a given acceptor and donor pair. The Förster distance $R_0$ is determined by the fluorescence quantum yield of the donor, the refractive index of the medium, the relative donor-acceptor orientation and the spectral overlap between the emission spectrum of the donor and the absorbance spectrum of the acceptor [1].

Since resonant Förster energy transfer takes place on short distances, the transfer involves the local density of states $\rho(\Delta\omega, \vec{r})$ [28] over a range of frequencies $\Delta\omega$ that is wide compared to the emission frequency $\omega_{Rad}$ of the donor: $\Delta\omega >> \omega_{Rad}$, see e.g. [29-31]. Consequently, the Förster transfer rate $\gamma_T$ is hardly affected by a change in local density of states at the donor emission frequency $\rho(\omega_{Rad}, \vec{r})$. We propose to use this property to tune the competition between resonant energy transfer and spontaneous emission.

## IV. TUNING EMISSION VERSUS ENERGY TRANSFER

The total decay rate $\gamma_{Tot}$ of a donor molecule is equal to the sum of its spontaneous emission rate $\gamma_{Rad}$, its Förster energy transfer rate $\gamma_T$ and its nonradiative decay rate $\gamma_{NR}$.

$$\gamma_{Tot} = \gamma_{Rad}\left(\rho(\omega_{Rad}, \vec{r})\right) + \gamma_T\left(\rho(\Delta\omega, \vec{r})\right) + \gamma_{NR} \tag{3}.$$

The spontaneous emission rate $\gamma_{Rad,}$ and the Förster energy transfer rate $\gamma_T$ depend on the local density of states as outlined above, whereas the nonradiative decay rate $\gamma_{NR}$ is independent of the local density of states. The quantum efficiency of the spontaneous emission of the donor $Q_{Rad}$ is equal to:

$$Q_{Rad} = \frac{\gamma_{Rad}\big(\rho(\omega_{Rad},\vec{r})\big)}{\gamma_{Rad}\big(\rho(\omega_{Rad},\vec{r})\big) + \gamma_T\big(\rho(\Delta\omega,\vec{r})\big) + \gamma_{NR}} \qquad (4)$$

and the quantum efficiency $Q_T$ for the transfer from the donor to the acceptor is equal to:

$$Q_T = \frac{\gamma_T\big(\rho(\Delta\omega,\vec{r})\big)}{\gamma_T\big(\rho(\Delta\omega,\vec{r})\big) + \gamma_{Rad}\big(\rho(\omega_{Rad},\vec{r})\big) + \gamma_{NR}} \qquad (5).$$

By changing the local density of states at the donor emission frequency $\rho(\omega_{Rad},\vec{r})$ (see Fig. 1B), the donor spontaneous emission rate $\gamma_{Rad}\big(\rho(\omega_{Rad},\vec{r})\big)$ is influenced, while leaving the energy transfer rate $\gamma_T\big(\rho(\Delta\omega,\vec{r})\big)$ and the nonradiative decay rate $\gamma_{NR}$ unaffected. Hence, in an energy transfer coupled system manipulation of $\rho(\omega_{Rad},\vec{r})$ not only results in a change in the total donor emission rate $\gamma_{Tot}$, but also in anticorrelated changes in the quantum efficiencies $Q_T$ and $Q_{Rad}$ of the two competing processes energy transfer and spontaneous emission. An increase in $\gamma_{Rad}\big(\rho(\omega_{Rad},\vec{r})\big)$ yields a higher quantum efficiency $Q_{Rad}$ of the spontaneous emission and a lower quantum efficiency $Q_T$ of the energy transfer. A decrease of the local density of states at the donor emission leads to the opposite effect, namely a lower quantum efficiency $Q_{Rad}$ of the spontaneous emission and a higher quantum efficiency $Q_T$ of the energy transfer.

The quantum efficiencies $Q_T$ and $Q_{Rad}$ are directly associated with the donor and acceptor emission intensities of the components of the coupled system. Thus, modulation of the local density of states at the donor emission frequency and consequent manipulation of the spontaneous donor emission rate $\gamma_{Rad}\left(\rho(\omega_{Rad},\vec{r})\right)$ results in a change of the intensity ratio of the donor and acceptor emission. This holds for continuous as well as pulsed excitation. Therefore an increase in the local density of states $\rho(\omega_{Rad},\vec{r})$ results in an increased relative donor intensity and decreased acceptor emission intensity. A decrease of the local density of states at the donor emission frequency leads to the opposite effect, namely to decreased relative donor emission intensity and increased acceptor emission intensity.

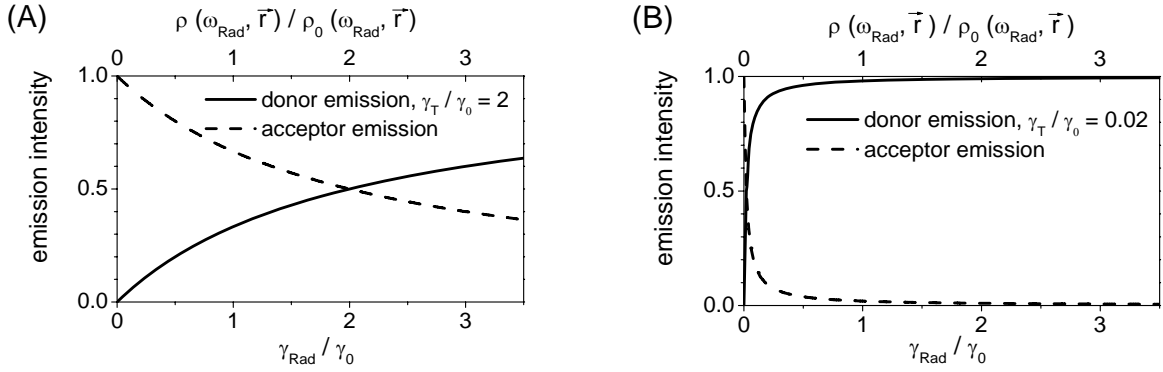

Figure 2: Dependence of donor and acceptor emission intensities as a function of the donor spontaneous emission rate $\gamma_{Rad}$ and the local density of states $\rho(\omega_{Rad},\vec{r})$ at the emission frequency of the donor; $\rho(\omega_{Rad},\vec{r})$ is normalized to the density of states in homogenous medium $\rho_0(\omega_{Rad},\vec{r})$, $\gamma_{Rad}$ and $\gamma_T$ are normalized to the donor emission rate in homogenous medium $\gamma_0$. (A) Effective energy transfer system, with $\gamma_T/\gamma_0 = 2$. Manipulating $\gamma_{Rad}/\gamma_0$ in a practical range between 0.5 and 2 leads to a very pronounced increase of donor emission and a concerted decrease of acceptor emission. (B) Weak energy transfer system, $\gamma_T/\gamma_0 = 0.02 << 1$. Manipulating $\gamma_{Rad}/\gamma_0$ between 0.5 and 2 leads to hardly any increase of donor emission or to decrease of acceptor emission. The evolution of the donor and acceptor emission intensities in dependence from $\gamma_{Rad}$ or $\rho(\omega_{Rad},\vec{r})$ shows the clear distinction between presence and absence of effective energy transfer.

In Figure 2 we present the expected change of emission intensities of the donor and acceptor intensities for effective and non-effective energy transfer. Since the energy transfer donor molecules in coupled systems are usually very efficient emitters [6], we neglect the nonradiative decay rate $\gamma_{NR}$ for simplicity. As illustrated in Figure 2A the expected effect is especially pronounced for cases where effective energy transfer occurs. Further, in the extreme limit of $\rho \rightarrow 0$ (corresponding to a photonic band gap), and for unity quantum efficiency, it is even possible to obtain the resonant energy transfer rate, since the donor's finite decay rate equals the transfer rate. It is interesting to note that in contrast an isolated high quantum efficiency dye molecule in the absence of energy transfer reveals no changes in the emission intensities when manipulating $\gamma_{Rad}$, see ref. [32].

The manipulation of the energy transfer donor density of states permits the analysis of Förster coupled systems that are difficult to address with conventional methods. The conventional approach to analyze Förster coupled systems (see [1]) is the separate characterization of the individual donor and acceptor chromophores and comparison with the spectral behavior of the coupled system. This method can not be applied for a number of biological systems because it is impossible to separate the components without unwanted alterations. We envision that tetrameric fluorescent proteins are excellent candidates to apply our approach. The most prominent member of this class of emitters is DsRed, a protein discovered in a reef coral [33].

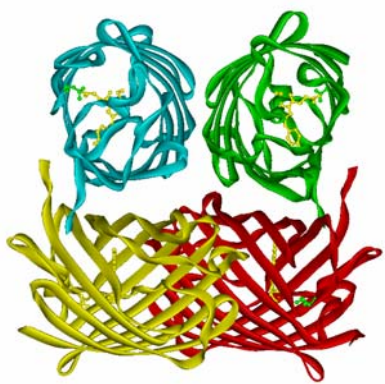

Figure 3: Structure of tetrameric red emitting proteins (color online). The fluorescent protein DsRed and its variants form tetramers even at low nanomolar concentrations. Each of the four monomers forming the tetramer contains either one green or one red emitting chromophore. The analysis of Förster coupling in biological systems like these protein tetramers is very difficult with conventional methods, since the monomers containing a green or a red emitting chromophore are always formed together and it is not possible to separate the proteins and obtain functional monomers.

DsRed forms tetramers even at exceedingly low nanomolar concentrations [34-37] (see Figure 3). Each protein monomer can contain either one green or one red emitting chromophore [38]. The assembly of these two different chromophores in a highly symmetric tetramer suggests a coupling of the chromophores by fluorescence resonance energy transfer. However, the separation of the monomers into functional proteins containing solely green or red emitting chromophores is impossible. Thus it is impossible to compare the spectral behavior of the individual donor and acceptor chromophores with the spectral behavior of the coupled system. Consequently the issue of energy transfer between unlike chromophores in DsRed tetramers remains a topic of continued discussion in the literature, and inconsistencies in the results reported reflect the intrinsic difficulties [34,39-42]. If the different chromophores in one DsRed tetramer indeed form an energy transfer coupled system, we expect the transfer efficiency to be high, which makes DsRed very suitable for tuning the competition between Förster transfer and spontaneous emission since the observable intensity changes of donor and acceptor emission is expected to be large (see Figure 2A). Conversely, if no or little energy transfer is taking place (Figure 2B), only the lifetime of the donor is affected by changing the local density of states, whereas the acceptor emission intensity remains completely unaffected.

We expect that this experimental situation is clearly distinguishable from the case where energy transfer is taking place within a protein tetramer, without the need to "take apart" a fluorescing protein tetramer.

## V. SUMMARY

In conclusion, we have theoretically shown the possibility to control the competition between spontaneous emission and energy transfer in Förster coupled systems by manipulation of the density of photonic states at the donor emission frequency. To achieve the control over the density of photonic states, the Förster coupled system can be placed near a metallic mirror, a dielectric interface or inside a photonic crystal. Our approach is especially suited for biological systems, since it allows the manipulation of Förster coupled systems in a continuous and nondestructive way. We expect that the manipulation of the local density of states can be a suitable tool to gain insights into delicate biological energy transfer coupled systems like complex fluorescent protein oligomers and light harvesting complexes.

In addition to the biophysical systems proposed here, we anticipate the use of our approach in other fields of research, where coupled light emitters cannot easily be physically separated. In general, the knowledge of and control over energy transfer in competition with spontaneous emission may be of importance for completely different applications, such as commercial lighting.

## VI. ACKNOWLEDGMENTS

We are grateful to Femius Koenderink for providing the density of states calculations in photonic crystals. This work is a part of the research program of the Stichting voor Fundamenteel Onderzoek der Materie (FOM) that is financially supported by the Nederlandse Organisatie voor Wetenschappelijk Onderzoek (NWO). WLV acknowledges an NWO-Vici fellowship and STW-NanoNed support.

## VII. REFERENCES.